\documentclass[11pt]{article}
\usepackage{erice,epsfig}

\def\00186{\em Matters Grav.}

\def\be{\begin{equation}}
\def\ee{\end{equation}}
\def\bea{\begin{eqnarray}}
\def\eea{\end{eqnarray}}

\newcommand{\lambdabar}{{\hbox{$\lambda_e$\kern-1.9ex\raise+0.45ex\hbox{--}
\kern+0.2ex}}}

\begin{document}
\vspace*{3cm}
\title{
\vspace*{-8ex}
{\normalsize\rightline{\rm }\rightline{\rm }}
\vspace*{4ex}
RELATIVISTIC HEAVY ION EXCITATION OF GIANT RESONANCES\footnote{Invited talk at the 
{\em Workshop on Electromagnetic Probes of Fundamental Physics}, Erice, Italy, October 2001.}}

\author{ C.A. BERTULANI }

\address{NSCL, Michigan State University, \\
East Lansing, MI 48824-1321, USA}

\maketitle\abstracts{
Giant resonances and giant resonances built 
on other giant resonances in nuclei are observed with very large cross sections in 
relativistic heavy ion collisions. A theoretical effort is underway to 
understand the reaction mechanism which leads to this process, as 
well as a better understanding of the microscopic properties
of multiphonon states, e.g., their strength, energy 
centroids, widths and anharmonicities.  
}

\section{Giant Resonances}

\subsection{Single giant resonances}

\label{subsec:prod} Giant resonances in nuclei were first observed in 1937
by Bothe and Gentner \cite{[1]} who obtained an unexpectedly large
absorption of 17.6 MeV photons (from the $^{7}$Li(p,$\gamma $) reaction) in
some targets. These observations were later confirmed by Baldwin and Klaiber
(1947) with photons from a betatron. In 1948, Goldhaber and Teller \cite{[2]}
interpreted these resonances (called isovector giant dipole resonances
(IVGDP)) with a hydrodynamical model in which rigid proton and neutron
fluids vibrate against each other, the restoring force resulting from the
surface energy. Steinwendel and Jensen \cite{[3]} later developed the model,
considering compressible neutron and proton fluids vibrating in opposite
phase in a common fixed sphere, the restoring force resulting from the
volume symmetry energy. The standard microscopic basis for the description
of giant resonances is the random phase approximation (RPA) in which giant
resonances appear as coherent superpositions of one-particle one-hole ($1p1h$%
) excitations in closed shell nuclei or two quasiparticle excitations in
open shell nuclei (for a review of these techniques, see, for example, ref. 
\cite{[4]}). The isoscalar quadrupole resonances were discovered in
inelastic electron scattering by Pitthan and Walcher (1971) and in proton
scattering by Lewis and Bertrand (1972). Giant monopole resonances were
found later and their properties are closely related to the compression
modulus of nuclear matter. Following these, other resonances of higher
multipolarities and giant magnetic resonances were investigated. Typical
probes for giant resonance studies are (a) $\gamma $'s and electrons for the
excitation of IVGDR, (b) $\alpha $-particles and electrons for the
excitation of isoscalar giant monopole resonance (ISGMR) and giant
quadrupole resonance (ISGQR), and (c) (p, n), or ($^{3}$He, t), for
Gamow-Teller resonances, respectively.

\begin{figure}
\begin{center}
\psfig{figure=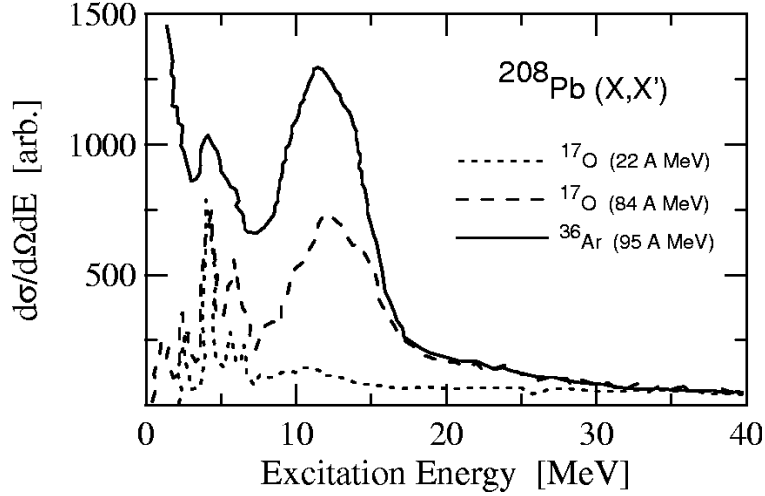,height=6.5cm,clip=}
\caption[...]{Experimental cross sections in arbitrary units for the excitation
of $^{208}$Pb targets by $^{17}$O (22.A MeV and 84.A MeV) and by $^{36}$Ar
(95.A MeV), as a function of the excitation energy.
\label{fig:xfel_princ}}
\end{center}
\end{figure}

\subsection{Multiphonon resonances}

\label{subsec:prod2} Inelastic scattering studies with heavy ion beams have
opened new possibilities in the field (for a review of the recent
developments, see ref. \cite{[5]}). A striking feature was observed when
either the beam energy was increased, or heavier projectiles were used, or
both \cite{[6]}. This is displayed in figure 1, where the excitation of the
GDR in $^{208}$Pb was observed in the inelastic scattering of $^{17}$O at 22
MeV/nucleon and 84 MeV/nucleon, respectively, and $^{36}$Ar at 95
MeV/nucleon \cite{[7],[8]}. What one clearly sees is that the `bump'
corresponding to the GDR at 13.5 MeV is appreciably enhanced. This feature
is solely due to one agent: the electromagnetic interaction between the
nuclei. This interaction is more effective at higher energies, and for
increasing charge of the projectile. In ref. \cite{[9]} it was shown that
the excitation probabilities of the GDR in heavy ion collisions approach
unity at grazing impact parameters. It was also obtained that, if the DGDR
(double GDR), or GDR$^{2}$ (i.e., a GDR excited on a GDR state), exists then
their excitation cross sections  in heavy ion collisions at
relativistic energies are of the order of hundreds of millibarns. The
calculation was based on the semiclassical approach, appropriate for heavy
ion scattering at high incident energies, and the harmonic oscillator model
for the giant resonances. The semiclassical model treats the relative motion
between the nuclei classically while quantum mechanics is used for the
internal degrees of freedom. In the harmonic picture for the internal
degrees of freedom the GDR is the first excited state in a harmonic well,
the DGDR is the second state, and so on. In ref. \cite{[10]} it was shown
that the excitation probabilities and cross sections are directly
proportional to the photonuclear cross sections for a given electric (E) and
magnetic (M) multipolarity. For an impact parameter b, excitation energy E ,
and a multipolarity $\pi \lambda $ ($\pi $ = E or M,\ $\lambda $ = 1, 2, 3, $%
\cdots $) the excitation probabilities are given by 
\begin{equation}
P_{\pi \lambda }(E,b)={\frac{1}{E}}N_{\pi \lambda }(E,b)\sigma _{\gamma
}^{\pi \lambda }(E)  \label{eq1}
\end{equation}
where $\sigma _{\gamma }^{\pi \lambda }(E)$ is the photonuclear cross
sections for the photon E and multipolarity $\pi \lambda $. The total
photonuclear cross section is $\sigma _{\gamma }(E)=\sum_{\pi \lambda
}\sigma _{\gamma }^{\pi \lambda }(E)$. In the semiclassical approach, the 
\textit{equivalent photon numbers} $N_{\pi \lambda }(E,b)$ are given
analytically \cite{[10]}. A quantum mechanical derivation of the excitation
amplitudes in relativistic Coulomb excitation shows that equation (1) can
also be obtained by using the saddle-point approximation in the DWBA
integrals \cite{[11]}. The total Coulomb excitation cross sections are then
obtained by an integration of equation (\ref{eq1}) over the impact parameter
b, including a factor, $T(b)$, which accounts for the strong absorption at
small impact parameters: $\sigma _{\pi \lambda }(E)=2\pi \int db\
b\;T(b)P_{\pi \lambda }(E,b)$. The total number of equivalent photons  
$n_{\pi \lambda}(E)=2\pi \int db\ b\;N_{\pi \lambda }(E,b)$ is 
given in \cite{[10],[11]}.
The cross section for the excitation of a giant resonance is obtained from
these expressions, by using the experimental photonuclear absorption cross
section for $\sigma _{\gamma }^{\pi \lambda }(E)$ in equation (\ref{eq1}).
One problem with this procedure is that the experimental photonuclear cross
section includes all multipolarities with the same weight: $\sigma _{\gamma
}^{exp}(E)=\sum_{\pi \lambda }\sigma _{\gamma }^{\pi \lambda }(E)$, while
the calculation based on equation (1) needs the isolation of $\sigma
_{\gamma }^{\pi \lambda }(E)$. This can only be done marginally, except in
some exclusive measurements. Generally, one finds in the literature the ($%
\gamma $, n), ($\gamma $, 2n), and ($\gamma $, 3n) cross sections, which
include the contribution of all multipolarities in the giant resonance
energy region. A separation of the different multipolarities can be obtained
roughly by use of sum rules, or some theoretical model for the nuclear
response to a photoexcitation. Assuming that one has $\sigma _{\gamma }^{\pi
\lambda }(E)$ somehow (either from experiments, or from theory), a simple
harmonic model based on the Axel--Brink hypotheses can be formulated to
obtain the probability to access a multiphonon state of order $n$. In the
harmonic oscillator model the inclusion of the coupling between all
multiphonon states can be performed analytically \cite{[9]}. One of the
basic changes is that the excitation probabilities calculated to
first-order, $P_{\pi \lambda }^{1st}(E,b)$, are modified to include the flux
of probability to the other states. That is, for the first harmonic state, 
\begin{equation}
P_{\pi \lambda }(E,b)=P_{\pi \lambda }^{1st}(E,b)\exp \left\{ -P_{\pi
\lambda }^{1st}(b)\right\} ,  \label{eq2}
\end{equation}
where $P_{\pi \lambda }^{1st}(b)$ is the integral of $P_{\pi \lambda
}^{1st}(E,b)$\ over the excitation energy $E$. In general, the probability
to reach a multiphonon state with the energy $E^{(n)}$ from the ground
state, with energy $E^{(0)}$, is obtained by an integral over all
intermediate energies 
\begin{eqnarray}
P_{\pi ^{\ast }\lambda ^{\ast }}^{(n)}(E^{(n)},b) &=&{\frac{1}{n!}}\exp
\left\{ -P_{\pi \lambda }^{1st}(b)\right\} \int dE^{(n-1)}dE^{(n-2)}\cdots
dE^{(1)}  \nonumber \\
&\times &P_{\pi \lambda }^{1st}(E^{(n)}-E^{(n-1)},b)P_{\pi \lambda
}^{1st}(E^{(n-1)}-E^{(n-2)},b)\cdots P_{\pi \lambda
}^{1st}(E^{(1)}-E^{(0)},b).  \label{eq3}
\end{eqnarray}
The character and spin assignment of the multipolarity $\lambda ^{\ast }$
depends on how the intermediate states couple with the electromagnetic
transition operators. For example, in the case of the DGDR (GDR$^{2}$ ),
assuming a $0^{+}$ ground state and excluding isospin impurities, the final
state has either spin and parity $0^{+}$ or $2^{+}$, respectively. A simpler
reaction model than the one above can be obtained by assuming that all
states can be approximated by a single isolated state. For example, we can
assume that the photoabsorption cross sections in the range of the GDR is
due to a single state with energy equal to the centroid energy of the GDR
exhausting the whole excitation strength. Then the multiphonon states are
sharp and equidistant. Eq. (3) becomes 
\begin{equation}
P_{\pi ^{\ast }\lambda ^{\ast }}^{(n)}(b)={\frac{1}{n!}}[P_{\pi \lambda
}^{1st}(b)]^{n}\exp \left\{ -P_{\pi \lambda }^{1st}(b)\right\} .  \label{eq4}
\end{equation}
The above relation was used to calculate the cross sections for the
excitation of the GDR, GDR$^{2}$, GDR$^{3}$, ISGQR and IVGQR in $^{136}$Xe,
respectively, for collisions with Pb nuclei as a function of the bombarding
energy, as shown in figure 2. Each resonance is considered to be a single
state exhausting 100\% of the respective sum rule. Also shown in the figure
is the geometrical cross section, (GC), $\sigma \sim \pi
(A_{P}^{1/3}+A_{T}^{1/3})^{2}$ fm$^{2}$ . The cross sections for the
excitation of the GDR$^{2}$ is large, of the order of hundreds of mb. Much
of the interest in looking for multiphonon resonances relies on the
possibility of finding exotic particle decay of these states. For example,
in ref. \cite{[15]} a hydrodynamical model was used to predict the proton
and neutron dynamical densities in a multiphonon state of a nucleus. Large
proton and neutron excesses at the surface are present in a multiphonon
state. Thus, the emission of exotic clusters from the decay of these states
are a natural possibility. A more classical point of view is that the
Lorentz contracted Coulomb field in a peripheral relativistic heavy ion
collision acts as a hammer on the protons of the nuclei \cite{[10]}. This
(collective) motion of the protons seems only to be probed in relativistic
Coulomb excitation. 

\begin{figure}[tbp]
\vskip0.5cm
\par
\begin{center}
\psfig{figure=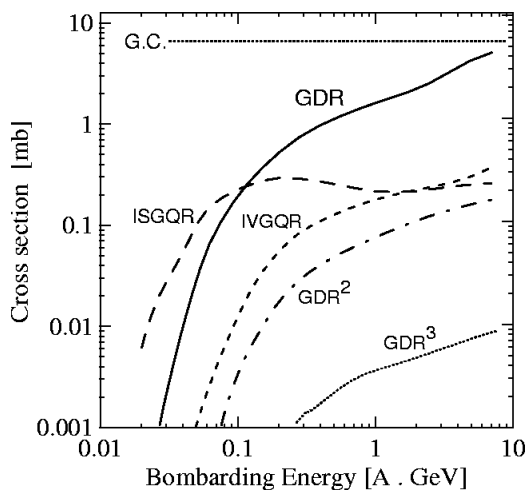,height=2.5in}
\end{center}
\caption{Theoretical cross sections for the excitation of the GDR, ISGQR,
IVGQR, GDR$^{2}$ and the GDR$^{3}$ , in the reaction $^{208}$Pb + $^{208}$%
Pb, as a function of the bombarding energy. }
\label{fig:fig2}
\end{figure}

Despite all the advantages of
relativistic Coulomb excitation, the DGDR was first found in pion scattering at the Los
Alamos Pion Facility \cite{[12]}. In pion scattering off nuclei the DGDR can
be described as a two-step mechanism induced by the pion-nucleus
interaction.
Only about five years later, the first Coulomb
excitation experiments for the excitation of the DGDR were performed at the
GSI facility in Darmstadt/Germany \cite{[13],[14]}. One of these
experiments observed  the neutron decay channels of giant resonances
excited by relativistic projectiles. The excitation spectrum of relativistic 
$^{136}$Xe projectiles incident on Pb were compared with the spectrum
obtained in C targets. A comparison of the two spectra immediately proved
that the nuclear contribution to the excitation is very small. Another
experiment \cite{[14]} dealt with the photon decay of the double giant
resonance. The advantages of relativistic Coulomb excitation of heavy ions
over other probes (pions, nuclear excitation, etc.) was clearly demonstrated
in several other GSI experiments \cite{[13],[14],[16],[17]}.

\section{Energy, width and strength of the DGDR}

Experimentally \cite{[18]}, it was found that the energy of the DGDR agrees
with the harmonic prediction, i.e., that its energy should be  twice
the energy of the GDR, although small departures from this prediction were
seen, especially in pion and nuclear excitation experiments. The width of
the DGDR seems to be $\sqrt{2}$ times that of
the GDR, although a value equal to 2 is not ruled out completely. An unexpected
result was obtained for the ratio between the measured
cross sections and the calculated ones. This seems to be strongly
dependent on the experimental probe. The largest values of $\sigma
_{exp}/\sigma _{th}$ come from pion experiments, yielding up to a value of 5
for it. In Coulomb excitation experiments this ratio was
initially \cite{[13],[14]} of the order of 2.

\subsection{Width of the DGDR}

In a microscopic approach, the GDR is described as a coherent superposition
of one-particle one-hole states. One of the many such states is pushed up by
the residual interaction to the experimentally observed position of the GDR.
This state carries practically all of the E1 strength. This situation is
simply realized in a model with a separable residual interaction. We write
the GDR state as (one phonon with angular momentum $1M$) $%
|1,1M>=A_{1M}^{\dagger }|0>$, where $A_{1M}^{\dagger }$ is a proper
superposition of particle-hole creation operators. Applying the quasi-boson
approximation we can use the boson commutation relations and construct the
multiphonon states (N-phonon states). An N-phonon state will be a coherent
superposition of N-particle N-hole states. The width of the GDR is
essentially due to the spreading width, i.e., to the coupling to more
complicated quasi-bound configurations. The escape width only plays a minor
role. Let us take a simple model for the strength function \cite{[19]}. A
state $|a>$ (i.e., a GDR state) is coupled by some mechanism to more
complicated states $|\alpha >$. For simplicity we assume a constant coupling
matrix element $V_{a\alpha }=<a|V|\alpha >=<\alpha |V|a>=v$. With an equal
spacing of D of the levels $|\alpha >$ one obtains a width $\Gamma =2\pi
v^{2}/D$ for the state $|a>$. We assume the same mechanism to be responsible
for the width of the N-phonon state: one of the N-independent phonons decays
into the more complicated states $|\alpha >$ while the other (N - 1)-phonons
remain spectators. We write the coupling interaction in terms of creation
(destruction) operators $c_{\alpha }^{\dagger }\ (c_{\alpha })$ of the
complicated states $|\alpha >$ as 
\begin{equation}
V=v(A_{1M}^{\dagger }c_{\alpha }+A_{1M}c_{\alpha }^{\dagger }).  \label{eq6}
\end{equation}
For the coupling matrix elements $v_{N}$, which connects an N-phonon state $%
|N>$ to the state $|N-1,\alpha >$ (N-1 spectator phonons) one obtains 
\begin{equation}
vN=<N-1,\alpha |V|N>=v<N-1|A_{1M}|N>=v\sqrt{N},  \label{eq7}
\end{equation}
i.e., one obtains for the width $\Gamma _{N}$ of the N-phonon state 
\begin{equation}
\Gamma _{N}=2\pi N{\frac{v^{2}}{D}}=N\Gamma \ .  \label{eq8}
\end{equation}
Thus, the factor N in (\ref{eq8}) arises naturally from the bosonic
character of the collective states. For the DGDR this would mean $\Gamma
_{2}=2\Gamma _{1}$. The data points seem to favor a lower multiplicative
factor.

We can also give a qualitative explanation for a smaller $\Gamma _{2}/\Gamma
_{1}$ value. First we note that the value $\Gamma _{2}/\Gamma _{1}=2$ can
also be obtained from a folding procedure, as given in equation (\ref{eq3}).
If the sequential excitations are described by Breit-Wigner (BW) functions $%
P_{\pi \lambda }(E)$ with the centroid $\mathcal{E}$ and the width $\Gamma $%
, the convolution (\ref{eq3}) yields a BW shape with the centroid at $2%
\mathcal{E}$ for the DGDR and the total width of $2\Gamma _{1}$ . However,
if one uses Gaussian functions (instead of BW) for the shape of one-phonon
states, it is easy to show that one also obtains a Gaussian for the N-phonon
shape, but with the width given by $\sqrt{N}\Gamma _{1}$. The latter
assumption seems inconsistent since the experimentalists use BW fits for the
shape of giant resonances, which are in good agreement with the experimental
data. However, one can easily understand that the result $\sqrt{N}\Gamma _{1}
$ is not restricted to a Gaussian fit. For an arbitrary sequence of two
excitation processes we have $<E>=<E_{1}+E_{2}>$ and $%
<E^{2}>=<(E_{1}+E_{2})^{2}>$; for uncorrelated steps it results in the
addition in quadrature $(\Delta E)^{2}=(\Delta E_{1})^{2}+(\Delta E_{2})^{2}$%
. Identifying these fluctuations with the widths up to a common factor, we
obtain for identical phonons $\Gamma _{2}=\sqrt{2}\Gamma _{1}$. The same
conclusion will be valid for any distribution function which, as the
Gaussian one, has a finite second moment, in contrast to the BW or Lorentzian
ones with second moment diverging. We may conclude that, in physical terms,
the difference between $\Gamma _{2}/\Gamma _{1}=2$ and $\Gamma _{2}/\Gamma
_{1}=\sqrt{2}$ is due to the different treatment of the wings of the
distribution functions which reflect small admixtures of remote states.

\begin{figure}[tbp]
\vskip0.5cm
\par
\begin{center}
\psfig{figure=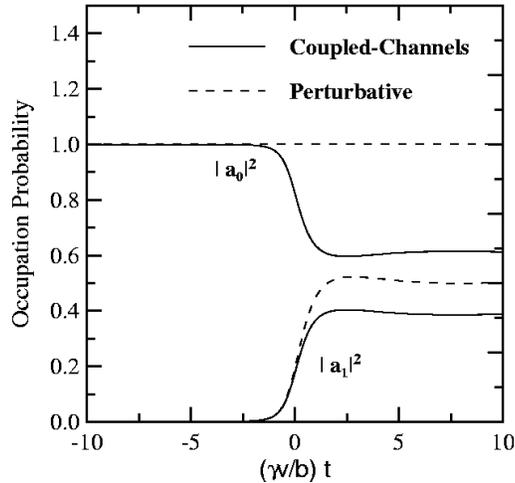,height=2.5in}
\end{center}
\caption{Occupation probability of the ground state, $|a_{0}|^{2}$, and of
the GDR, $|a_{1}|^{2}$, for the reaction $^{208}$Pb + $^{208}$Pb at 640.A
MeV, as a function of the reaction time. The reaction time is given in terms
of the adimensional quantity $\protect\gamma vt/b$, with b equal to the
impact parameter in the collision. The broken curves are the predictions of
perturbation theory, while the full curves are the predictions of
coupled--channels calculations. }
\label{fig:fig3}
\end{figure}

\subsection{Strength of the DGDR}

Microscopically, the harmonic picture is accomplished within the RPA
approximation. The excited states of the nucleus are described as
superpositions of particle-hole configurations with respect to the ground
state. The multiphonon resonances are built using products of the $1^{-}$
resonance states, yielding $0^{+}$ and $2^{+}$ double phonon states. The
interaction with the projectile is described in terms of a linear
combination of particle-hole operators weighted by the time-dependent field
for a given multipolarity of the interaction. Since the time-dependent
Coulomb field of a nucleus does not carry monopole multipolarity, the DGDR
states can be reached via two-step E1 transitions and direct E2 transitions
(for a $0^{+}$ ground state). Early calculations failed to explain the
experimental data.

There seems to be two possible reasons for $\sigma _{exp}/\sigma _{th}\neq 1$%
; (a) either the Coulomb excitation mechanism is not well described, or (b)
the response of the nucleus to two-phonon excitations is not well known.

Many authors studied the effects of the excitation mechanism of the
DGDR. In ref. \cite{[20]} the cross sections were
calculated using second-order perturbation theory. It was found that the
theoretical values were smaller than the experimental ones by a factor of
1.3 -- 2. However, it was suggested \cite{[21]} that second-order
perturbation theory is not adequate for relativistic Coulomb excitation of
giant resonances with heavy ions and that it is necessary to perform a
coupled channels calculation. We can see this more clearly in figure \ref
{fig:fig3}, taken from \cite{[22]}, where a coupled-channels study of
multiphonon excitation by the nuclear and Coulomb interactions in
relativistic heavy ion collisions was performed. The figure shows the
probability amplitude to excite the GDR in $^{208}$Pb, $|a_{1}|^{2}$, and
the occupation probability of the ground state, $|a_{0}|^{2}$, for a grazing
collision of $^{208}$Pb + $^{208}$Pb at 640 MeV/nucleon. The broken curves
are the predictions of the first-order perturbation theory. We see that the
asymptotic excitation probability of the GDR is quite large ($~$40\%). In
first-order perturbation theory the occupation probability of the ground
state is kept constant, equal to unity. Obviously, one greatly violates the
unitarity condition in this case. An  appropriate coupled-channels
calculation (full curves) shows that the ground-state occupation probability
has to decrease to meet the unitarity requirements, while the excitation
probability of the GDR is also reduced slightly for the same reason. In ref. 
\cite{[22]} it was shown that a good coupled-channels calculation does not
need to account for the exact coupling equations in all channels. The
strongest coupling, responsible for the effect observed in figure \ref
{fig:fig3} is the coupling between the ground state and the GDR states. This
has to be treated exactly within a coupled-channels calculation. The
coupling between the GDR and the other states (including the DGDR, IVGQR,
ISGQR, etc.) can be treated perturbatively.

The results of \cite{[22]} showed appreciable dependence of the excitation
cross sections of the DGDR on the width of both the GDR and the DGDR for $%
^{208}$Pb + $^{208}$Pb at 640 MeV/nucleon. It was also shown that the most
favorable energies for the measurement of the DGDR corresponds to the SIS
energies at the GSI/Darmstadt facility.

\begin{figure}[tbp]
\vskip0.5cm
\par
\begin{center}
\psfig{figure=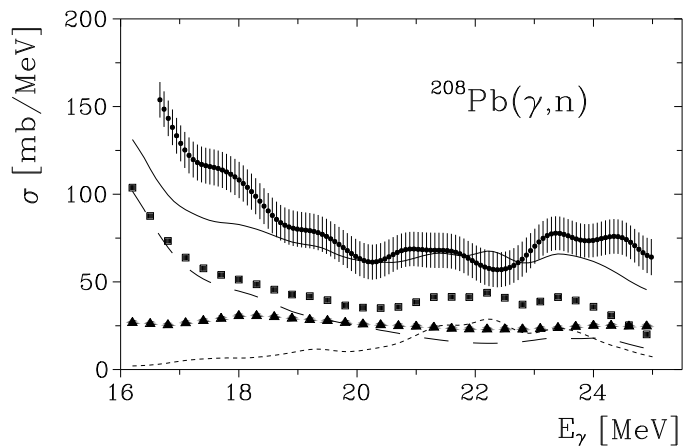,height=2.5in}
\end{center}
\caption{ Photoneutron cross section for $^{208}$Pb. Experimental data (dots
with experimental errors) are from ref. [28]. The long-broken curve is the
high-energy tail of the GDR, the short-broken curve is the IVGQR and the
curve with squares is their sum. The contribution of two-phonon states is
plotted by a curve with triangles. The full curve is the total calculated
cross section. }
\label{fig:fig4}
\end{figure}

\subsection{Anharmonicities}

Another possible effect arises from a shift of the energy centroid of the
DGDR due to anharmonic effects \cite{[23]}. In ref. \cite{[22]} one obtained 
$\sigma _{DGDR}$ = 620, 299 and 199 mb for the centroid energies of $E_{DGDR}
$ = 20, 24 and 27 MeV, respectively. This shows that anharmonic effects can
play a big role in the Coulomb excitation cross sections of the DGDR,
depending on the size of the shift of $E_{DGDR}$. However, in ref. \cite
{[20]} the source for anharmonic effects were discussed and it was suggested
that it should be very small, i.e., $\Delta ^{(2)}E=E_{DGDR}-2E_{GDR}\simeq 0
$. The anharmonic behavior of the giant resonances as a possibility to
explain the increase of the Coulomb excitation cross sections has been
studied by several authors \cite{[23],[24]} (see also ref. \cite{[25]}, and
references therein). It was found that the effect is indeed negligible and
it could be estimated \cite{[25]} as $\Delta ^{(2)}E<E_{GDR}/(50A)\sim
A^{-4/3}$ MeV. Recent studies \cite{daniel} of the reaction mechanism with
anharmonic effects  support the idea that these are indeed very small.

\subsection{Other routes to the DGDR}

From the above discussion we see that the magnitude of the Coulomb
excitation cross sections of the DGDR can be affected due to uncertainties
in: (a) strength, (b) width, (c) energies, or (d) reaction mechanism. Cases
(a) and (c) are the basis of the Axel-Brink hypothesis and we have seen that
a modification of their values would only be considered seriously if
anharmonic effects were large, which does not seem to be the case. Case
(b) is an open question. Microscopic calculations \cite{[24]} have shown
that, taking into account the Landau damping, the collective state splits
into a set of different $1_{i}^{-}$ states distributed over an energy
interval, where $i$ is the order number of each state. A further
fragmentation of the $1_{i}^{-}$ states into thousands of closed packed
states, is obtained by the coupling of one- and two-phonon states. This
leads to a good estimate of the spreading width of the GDR. However, the
DGDR states were obtained by a folding procedure: 
\begin{equation}
|[1_{i}^{-}\otimes 1_{i^{\prime }}^{-}]_{J^{\pi
}=0^{+},1^{+},2^{+}}>_{M}=\sum_{m,m^{\prime }}(1m1m^{\prime
}|JM)|1_{i}^{-}>_{m}|1_{i^{\prime }}^{-}>_{m^{\prime }}.
\end{equation}
\begin{figure}[tbp]
\vskip0.5cm
\par
\begin{center}
\psfig{figure=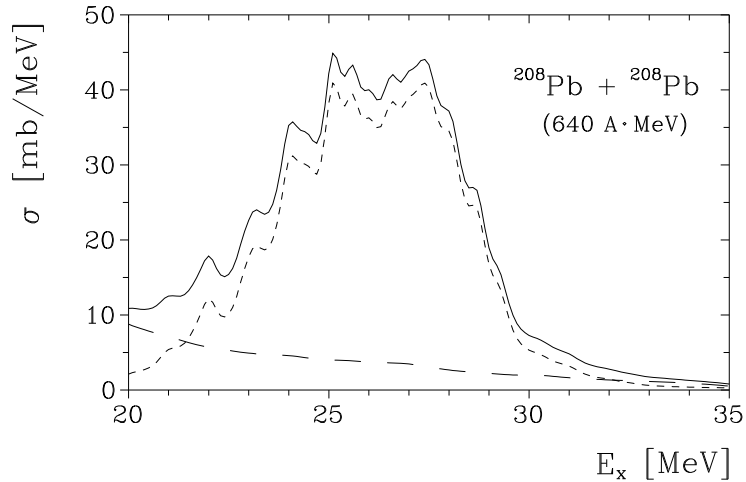,height=2.5in}
\end{center}
\caption{ The contribution for the excitation of two-phonon 1$^{-}$ states
(long-broken curve) in first-order perturbation theory, and of two-phonon 0$%
^{+}$ and 2$^{+}$ DGDR states to second-order (short-broken curve). The
total cross section (for $^{208}$Pb(640.A MeV) + $^{208}$Pb) is shown by the
full curve. }
\label{fig:fig8}
\end{figure}

The width of the DGDR is thus fixed by the width of the GDR. It is therefore
impossible to make any quantitative prediction for the width of the DGDR,
other than saying that $\sqrt{2}\leq \Gamma _{DGDR}/\Gamma _{GDR}\leq 2$. Thus, we
return to the discussion of the reaction mechanism, and how it could affect
the magnitude of the cross sections. The nuclear excitation of giant
resonances is very small in magnitude compared with Coulomb excitation in
collisions with heavy ions at relativistic energies \cite{[22]}. In ref. \cite
{[22]} it was shown that the nuclear-Coulomb interference  
is also a small effect.

In ref. \cite{[26]} the contribution of non-natural parity $1^{+}$
two-phonon states were investigated in a coupled-channels calculation. The
diagonal components $[1_{i}^{-}\otimes 1_{i}^{-}]_{1^{+}}$ are forbidden by
symmetry properties but non-diagonal ones $[1_{i}^{-}\otimes 1_{i^{\prime
}}^{-}]_{1^{+}}$ , may be excited in the two-step process bringing some
``extra strength'' in the DGDR region. Consequently, the role of these
non-diagonal components depends on how strong the Landau damping is. A
coupled-channels calculation found that the contribution of the 1$^{+}$
states to the total cross section is small. The reason for this is better
explained in second-order perturbation theory. For any route to a final
magnetic substate $M$, the second-order amplitude will be proportional to $%
(001\mu |1\mu )V_{E1\mu ,0\rightarrow 1^{-}}\times (1\mu 1\mu |1M)V_{E1\mu
,1^{-}\rightarrow 1^{+}}+(\mu \leftrightarrow \mu ^{\prime })$, where $%
V_{E1\mu ,i\rightarrow f}$ is the $\mu $-component of the interaction
potential (for a spin-zero ground state, $\mu $ is also the angular momentum
projection of the intermediate state). Assuming that the phases and the
products of the reduced matrix elements for the two sequential excitations
are equal, we obtain $V_{E1\mu ,0\rightarrow 1^{-}}\times V_{E1\mu ^{\prime
},1^{-}\rightarrow 1^{+}}=V_{E1%
{\mu}
^{\prime },0\rightarrow 1^{-}}\times V_{E1\mu ,1^{-}\rightarrow 1^{+}}$.
Thus, under these circumstances, and since $(001\mu |1\mu )=1$, we get an
identically zero result for the excitation amplitude of the $1^{+}$ DGDR
state as a consequence of $(1\mu 1\mu |1M)=-(1\mu ^{\prime }1\mu |1M)$. 

We note that multiphonon states can be obtained by coupling all kinds of
phonons. Each configuration $[\lambda _{1}^{\pi _{1}}\otimes \lambda
_{2}^{\pi _{2}}]$ can be obtained theoretically from a sum over several
two-phonon states made of phonons and of complicated states with a given
spin and parity $\lambda _{1}^{\pi _{1}}$, $\lambda _{2}^{\pi _{2}}$ , and
different RPA root numbers $i_{1}$ , $i_{2}$ of their constituents. The
cross sections can be obtained accordingly: 
\begin{equation}
\sigma ([\lambda _{1}^{\pi _{1}}\otimes \lambda _{2}^{\pi
_{2}}])=\sum_{i_{1},i_{2}}\sigma ([\lambda _{1}^{\pi _{1}}(i_{1})\otimes
\lambda _{2}^{\pi _{2}}(i_{2})]).  \label{eq12}
\end{equation}
As an example, in ref. \cite{[27]} the total number of two-phonon $1^{-}$
states generated in this way was about 10$^{5}$. The absolute value of the
photoexcitation of any two-phonon state under consideration is negligibly
small but altogether they produce a sizeable cross section. The $1^{-}$
two-phonon states obtained in ref. \cite{[27]} were used to calculate their
contribution to the ($\gamma $,n) cross section in $^{208}$Pb, via direct E1
excitations. This is shown in figure \ref{fig:fig4}. Experimental data (dots
with experimental errors) are from \cite{[28]}. The long-broken curve is the
high-energy tail of the GDR, the short-broken curve is the IVGQR and the
curve with squares is their sum. The contribution of two-phonon $1^{-}$
states is plotted by a curve with triangles. The full curve is the total
calculated cross section. Thus, already at the level of photonuclear data
the contribution of two-phonon $1^{-}$ states is of relevance. Here they are
not reached via two-step processes, but in direct excitations. Since the
energy region of these states overlap with that of the DGDR, in Coulomb
excitation experiments they should also contribute appreciably. In fact, it
was shown \cite{[27]} that their contribution to the total cross
section for $^{208}$Pb + $^{208}$Pb (640.A MeV) in the DGDR region is of the
order of 15\%. In figure \ref{fig:fig8} we show the contribution of the
excitation of two-phonon $1^{-}$ states (long-dashed curve) in first order
perturbation theory, and for two-phonon $0^{+}$ and $2^{+}$ DGDR states in
second-order (short-dashed curve). The total cross section (for $^{208}$Pb
(640.A MeV) + $^{208}$Pb) is shown by the solid curve.

\begin{figure}[tbp]
\vskip0.5cm
\par
\begin{center}
\psfig{figure=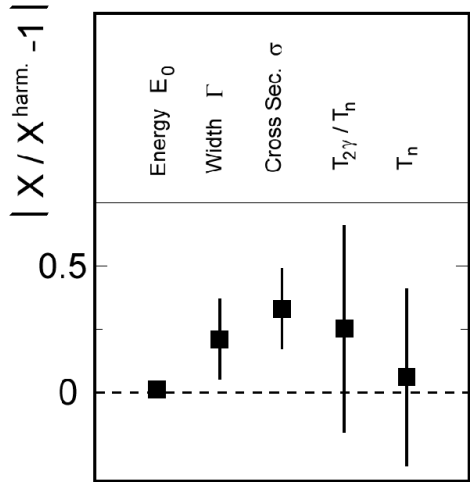,height=2.5in}
\end{center}
\caption{Deviation of the experimental results from the harmonic oscillator
prediction for the energy, width, cross section, ratio between the decay by
emission of two gammas, and of the one---neutron decay width, respectively. }
\label{fig:fig7}
\end{figure}

\section{Present situation and perspectives}

Presently, experiments tell us that the harmonic model reproduces the cross
section for the GDR quite well, but it gives smaller values than the measured cross
sections by as much as 30\%. In figure \ref{fig:fig7} the present situation on our
knowledge of the energy, width, excitation cross section, branching ratio
for gamma to neutron emission, and the neutron emission width, respectively,
is shown in comparison with calculations based on the simple harmonic
picture. We see that the theory-experiment agreement is much better than
that obtained in the pioneer experiments. As we have seen in this
short review there are several effects which compete in the excitation of
double giant resonances in relativistic Coulomb excitation. These effects
were discovered in part by the motivation to explain discrepancies between
the harmonic picture of the giant resonances and the recent experimental
data. We cannot say at the moment how much we have progressed towards a
better understanding of these nuclear structures, as some controversies
still remain in the literature (see, for example, \cite{[29]}). Recent
studies of giant resonances in ultra-relativistic
collisions have been performed at CERN \cite{[30]} and Brookhaven \cite{[31]}. 
Since the nuclei fragment after the excitation to a giant resonance, this
process can be used for beam monitoring as well \cite{[31]}. 
The field is just in its infancy and important experimental and
theoretical progress will occur in the near future.

\end{document}